\documentclass[aps,prl,reprint,superscriptaddress]{revtex4-2}
\usepackage{graphicx}
\usepackage{graphicx}
\usepackage{dcolumn}
\usepackage{bm}
\usepackage[dvipsnames]{xcolor}
\usepackage{amsmath}

\begin{document}

\title{Thermal cycling induced alteration of the stacking order and spin-flip in the room temperature van der Waals magnet Fe$_5$GeTe$_2$}

\author{Xiang Chen} 
\email{xiangchen@berbeley.edu}
\affiliation{Materials Science Division, Lawrence Berkeley National Lab, Berkeley, California 94720, USA}
\affiliation{Physics Department, University of California, Berkeley, California 94720, USA}

\author{Wei Tian} 
\affiliation{Neutron Scattering Division, Oak Ridge National Laboratory, Oak Ridge, Tennessee 37831, USA }

\author{Yu He}
\affiliation{Department of Applied Physics, Yale University, New Haven, Connecticut, 06511, USA}
\affiliation{Physics Department, University of California, Berkeley, California 94720, USA}
\affiliation{Materials Science Division, Lawrence Berkeley National Lab, Berkeley, California 94720, USA}

\author{Hongrui Zhang}
\affiliation{Department of Materials Science and Engineering, University of California, Berkeley, California 94720, USA}

\author{Tyler L. Werner}
\affiliation{Department of Applied Physics, Yale University, New Haven, Connecticut, 06511, USA}

\author{Saul Lapidus} 
\affiliation{Advanced Photon Source, Argonne National Laboratory, Argonne, Illinois 60439, USA}

\author{Jacob P.C. Ruff}
\affiliation{Cornell High Energy Synchrotron Source, Cornell University, Ithaca, NY 14853, USA}

\author{Ramamoorthy Ramesh}
\affiliation{Department of Materials Science and Engineering, University of California, Berkeley, California 94720, USA}
\affiliation{Materials Science Division, Lawrence Berkeley National Lab, Berkeley, California 94720, USA}
\affiliation{Physics Department, University of California, Berkeley, California 94720, USA}

\author{Robert J. Birgeneau}
\affiliation{Physics Department, University of California, Berkeley, California 94720, USA}
\affiliation{Materials Science Division, Lawrence Berkeley National Lab, Berkeley, California 94720, USA}
\affiliation{Department of Materials Science and Engineering, University of California, Berkeley, California 94720, USA}

\date{\today}

\begin{abstract}

The magnetic properties of the quasi-two-dimensional van der Waals magnet Fe$_{5-\delta}$GeTe$_2$ (F5GT), which has a high ferromagnetic ordering temperature $T_{\text{C}}$ $\sim$ 315 K, remains to be better understood. It has been demonstrated that the magnetization of F5GT is sensitive to both the Fe deficiency ${\delta}$ and the thermal cycling history. Here, we investigate the structural and magnetic properties of F5GT with a minimal Fe deficiency ($|{\delta}|$ $\le$ 0.1), utilizing combined x-ray and neutron scattering techniques. Our study reveals that the quenched F5GT single crystals experience an irreversible, first-order transition at $T_{\text{S}}$ $\sim$ 110 K upon first cooling, where the stacking order partly or entirely converts from ABC-stacking to AA-stacking order. Importantly, the magnetic properties, including the magnetic moment direction and the enhanced $T_{\text{C}}$ after the thermal cycling, are intimately related to the alteration of the stacking order. Our work highlights the significant influence of the lattice symmetry to the magnetism in F5GT.

\end{abstract}

\maketitle

\section{Introduction}

The quasi-two-dimensional (quasi-2D) van der Waals (vdW) magnets are prominent systems for studying low dimensional magnetism and anomalous transport behaviors \cite{Park_2016_JPCM, Burch_2018_Nature, Gibertini_2019_NatNano, Gong_2019_Science, Mak_2019_NatRevPhy, Sierra_2021_NatNano, Gong_2017_Nature, Huang_2017_Nature}. Among them, the itinerant Fe$_{5-\delta}$GeTe$_2$ (F5GT) system is particularly interesting because of its advantageous characteristics for potential room-temperature (RT) spintronic applications \cite{Stahl_2018_ZFAAC_F5GT, May_2019_ACSnano_F5GT, Zhang_2020_PRB_F5GT, May_2019_PRM, Yang_2021_PRB}. The bulk form of F5GT has a near RT ferromagnetic (FM) transition temperature $T_{\text{C}}$ $\approx$ 275 K $\sim$ 330 K \cite{Stahl_2018_ZFAAC_F5GT, May_2019_ACSnano_F5GT, Zhang_2020_PRB_F5GT, May_2019_PRM, Ly_2021_AFM, Gao_2020_AdvMat, Alahmed_2021_2DM, Ribeiro_2022_2DMA, Nair_2022_NanoRes, Chen_2022_2DM}. Owing to the coupling between the electronic and magnetic degrees of freedom, and the tunable sample thickness, the magnetic properties of F5GT are highly susceptible to perturbations \cite{Yamagami_2021_PRB, Wu_2021_PRB, Yamagami_2022_PRB, Huang_2022_CPB, May_2020_PRM_FCGT, Tian_2020_APL_FCGT, Zhang_2021_FCGT, Zhang_2022_FCGT_sciadv, Chen_2022_PRL, Tan_2021_NanoLett, Li_2022_AdvMat, Gao_2022_PRB}. For example, enhanced FM transition temperatures, from RT to 400 K and beyond, can be achieved by replacing Fe with either cobalt (Co) or nickel (Ni) \cite{May_2020_PRM_FCGT, Tian_2020_APL_FCGT, Zhang_2021_FCGT, Zhang_2022_FCGT_sciadv, Chen_2022_PRL}, or through applying high pressure \cite{Li_2022_AdvMat}. Also, a magnetic phase transition from a FM to antiferromagnetic (AFM) ground state can be obtained via Co doping or by voltage-induced charge doping \cite{May_2020_PRM_FCGT, Zhang_2021_FCGT, Tan_2021_NanoLett}. Interestingly, topological spin textures have been realized via engineering the lattice symmetry by Co doping or tuning the sample thickness \cite{Zhang_2021_FCGT, Zhang_2022_FCGT_sciadv, Gao_2022_PRB}. The aforementioned observations highlight the immense capacity for exploring both novel physical phenomena and technological applications in F5GT.

Despite the large amount of research directed at characterizing the magnetism in F5GT, its magnetic ground state remains contested \cite{May_2019_ACSnano_F5GT, May_2019_PRM, Zhang_2020_PRB_F5GT, Ly_2021_AFM, Alahmed_2021_2DM, Li_2022_AdvMat}. Specifically, conflicting suggestions about the spin structure have been reported, ranging from out-of-plane FM \cite{May_2019_ACSnano_F5GT, May_2019_PRM}, to in-plane FM \cite{Zhang_2020_PRB_F5GT}, and to ferrimagnetic order \cite{Alahmed_2021_2DM}. This is partly due to the experimental techniques utilized for investigating the magnetic properties in F5GT which are either only surface sensitive or indirect probes of the microscopic magnetic structure, such as the magnetic force microscopy (MFM), nitrogen-vacancy (NV) center magnetometry and magneto-transport measurements \cite{Zhang_2020_PRB_F5GT, Ly_2021_AFM, Li_2022_AdvMat, Chen_2022_2DM}. Further complexity is added because the magnetism of F5GT is sensitive to both the iron (Fe) deficiency (${\delta}$ $\ge$ 0), especially the half-occupied Fe1 site, and the growth conditions during the crystal synthesis \cite{May_2019_PRM}. Therefore, it is important to have a thorough investigation of the microscopic lattice and magnetic structure in F5GT. Here, we utilize combined x-ray and neutron scattering techniques to study the bulk properties of F5GT. Our work indicates that the quenched F5GT single crystals, which exhibit thermal-cycling history dependence according to the magnetization measurements, actually undergo a first-order structural transition at $T_{\text{S}}$ $\sim$ 110 K upon first cooling. During this irreversible process, the stacking order of the sample partly or completely converts from ABC-stacking to AA-stacking order. Our neutron diffraction data indicate that the spin moment direction is directly controlled by the stacking order of the lattice, likely resulting from the strong magneto-elastic coupling and the interlayer exchange couplings. Our study highlights the significance of the stacking order and the interlayer couplings in determining the magnetic ground state of F5GT.

\begin{figure}[t]
\centering
\includegraphics[width= 8.5 cm]{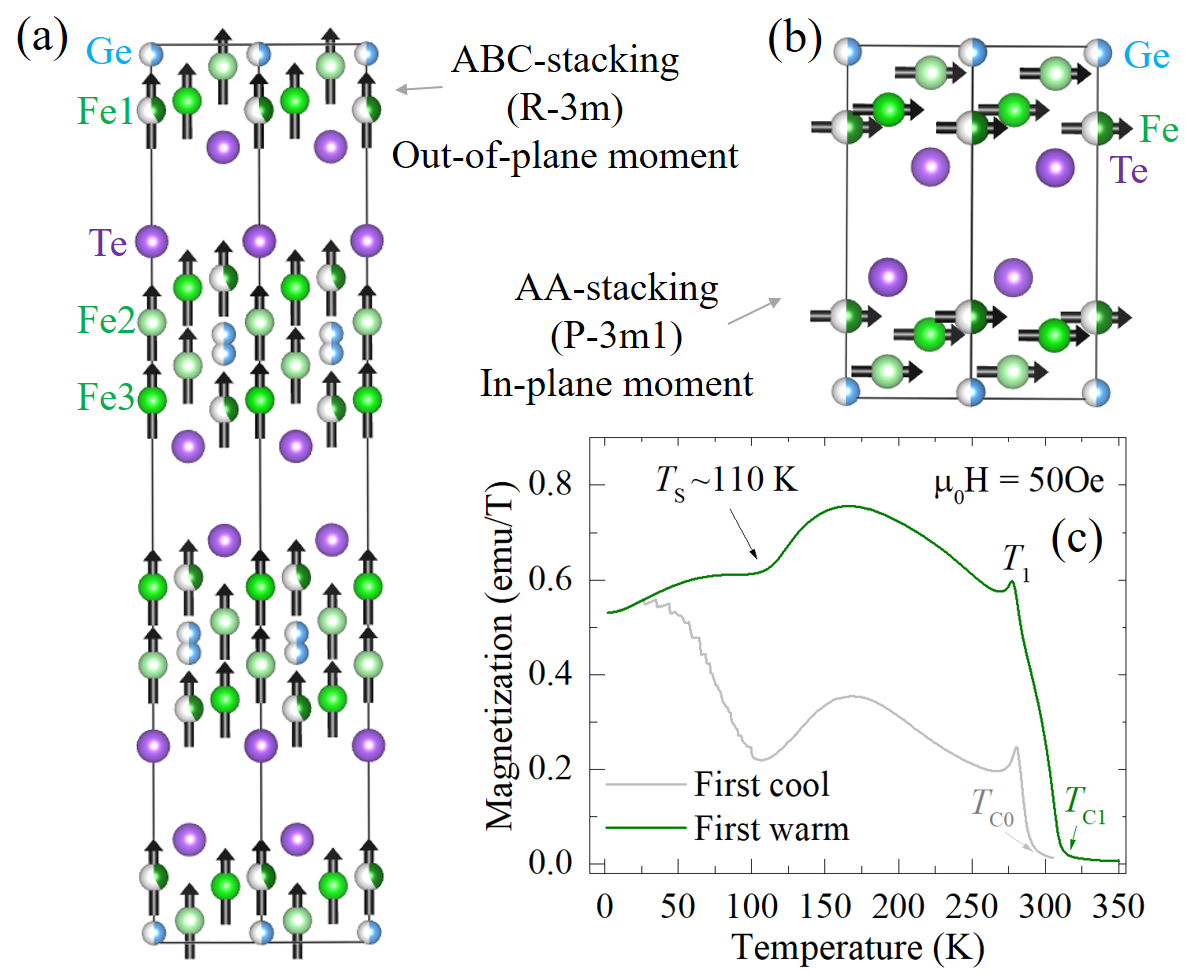}
\caption{(Color online) (a)-(b) The crystal structure of Fe$_5$GeTe$_2$ (F5GT) with two different types of stacking order: (a) ABC-stacking order with a space group (SG) R$\bar{3}$m and (b) AA-stacking order with a SG P$\bar{3}$m1 \cite{May_2020_PRM_FCGT, Zhang_2021_FCGT, Zhang_2022_FCGT_sciadv}. (c) Thermal cycling history dependence of the in-plane magnetization of the quenched F5GT single crystals. Prior to the first cooling process, the ferromagnetic onset temperature $T_{\text{C0}}$ is $\sim$ 285 K (gray line), which is enhanced to $T_{\text{C1}}$ $\sim$ 310 K after cooling below $T_{\text{S}}$ $\sim$ 110 K (green line).}
\label{fig:Fig1_alpha}
\end{figure}

\section{Results and Discussion}

Single crystals of F5GT were obtained via the chemical vapor transfer technique and quenched at 750 $^\circ$C to improve the sample quality during the last step of the synthesis \cite{May_2019_ACSnano_F5GT, May_2019_PRM, Zhang_2021_FCGT, Zhang_2022_FCGT_sciadv}. The cation deficiency ($|\delta| \le 0.1$) of the crystals, verified by energy dispersive x-ray spectroscopy, indicates that our samples are nearly stoichiometric. The x-ray scattering experiments were carried out at 44 keV at beamline QM2 of the Cornell High Energy Synchrotron Source (CHESS). A Pilatus 6M area detector was used to collect the diffraction pattern with the sample rotated 365 degrees around three different axes at 0.1$^\circ$ step and 0.1 s/frame data rate at each temperature. Elastic neutron scattering measurements were carried out using the 14.6-meV fixed-incident-energy triple-axis spectrometer HB-1A of the HFIR at ORNL, with a collimation of 40$'$-40$'$-40$'$-80$'$. For better comparison, the Bragg peaks $\textbf{Q}$ = $(H\cdot\frac{2\pi}{a}, K\cdot\frac{2\pi}{b}, L\cdot\frac{2\pi}{c})$ are defined in reciprocal lattice units ($r.l.u.$) with lattice parameters $a$ = $b$ $\approx$ 4.03 \AA \, and $c$ $\approx$ 29.2 \AA \, for both ABC-stacking and AA-stacking ordered domains of F5GT (Figs. 1(a)-(b)). The single crystals used for both the x-ray and neutron scattering experiments were freshly thermal-quenched prior to the measurements.

Consistent with the reported results \cite{May_2019_PRM}, the quenched F5GT single crystal exhibits a thermal history dependence as observed in the magnetization measurement (Fig. 1(c)). Prior to any cooling process, the F5GT sample has a magnetic onset temperature $T_{\text{C0}}$ $\sim$ 285 K (Fig. 1(c), gray line). Upon the first cooling process from RT down to temperature $T$ = 2 K (labelled as ``first cool"), the F5GT sample experiences an irreversible, first order structural transition at $T_{\text{S}}$ $\sim$ 110 K, as indicated by the large dip in the magnetization data (Fig. 1(c)). After this initial cooling process, the F5GT sample is warmed up (labelled as ``first warm" or ``thermally cycled") and has a permanently enhanced magnetic transition temperature $T_{\text{C1}}$ $\sim$ 310 K (Fig. 1(c), green line). Further thermal cycling (either cooling or warming) of the crystal, as long as it is below $\sim$ 550 K, does not affect the magnetization \cite{May_2019_PRM}.

\begin{figure}[t]
\centering
\includegraphics[width = 8.5 cm]{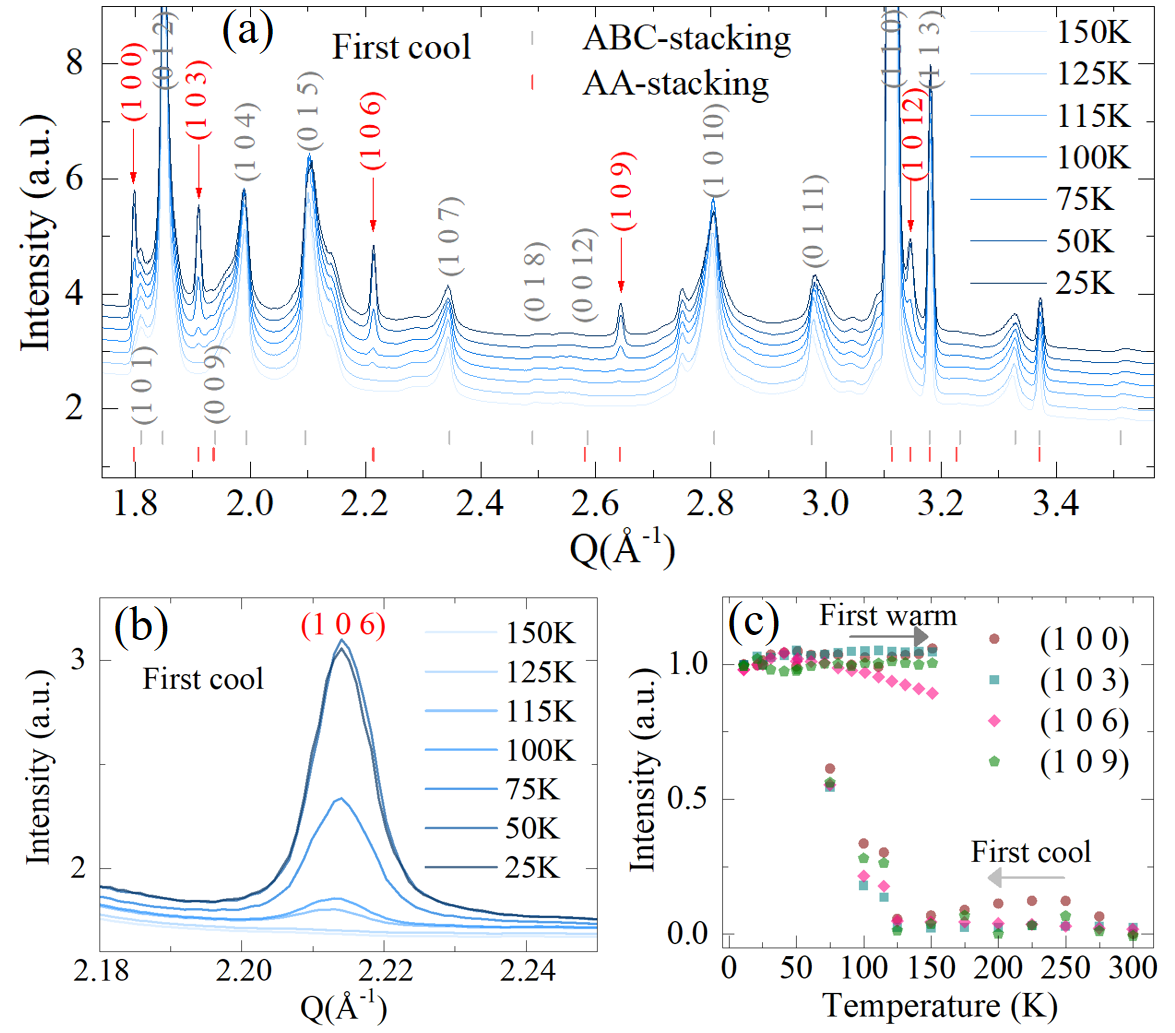}
\caption{(Color online) (a) Temperature dependence of the x-ray diffraction pattern of F5GT, during the ``first cool" procedure, collected with a decreasing temperature sequence. Vertical lines indicate the peak positions predicted by the two different models: ABC-stacking order with SG R$\bar{3}$m (gray) and AA-stacking with SG P$\bar{3}$m1 (red). Emerging peaks below $T_{\text{S}}$ $\sim$ 110 K are marked and labelled in red. (b) Magnified view of the (1, 0, 6) peak at select temperatures. (c) Temperature dependence of the normalized peak intensity at select peaks $\textbf{Q}_\text{AA}$ = (1, 0, 3$N$) ($N$ = 0, 1, 2, 3). }
\label{fig:Fig2_alpha}
\end{figure}

\begin{figure*}[t]
\centering
\includegraphics*[width= 14.5 cm]{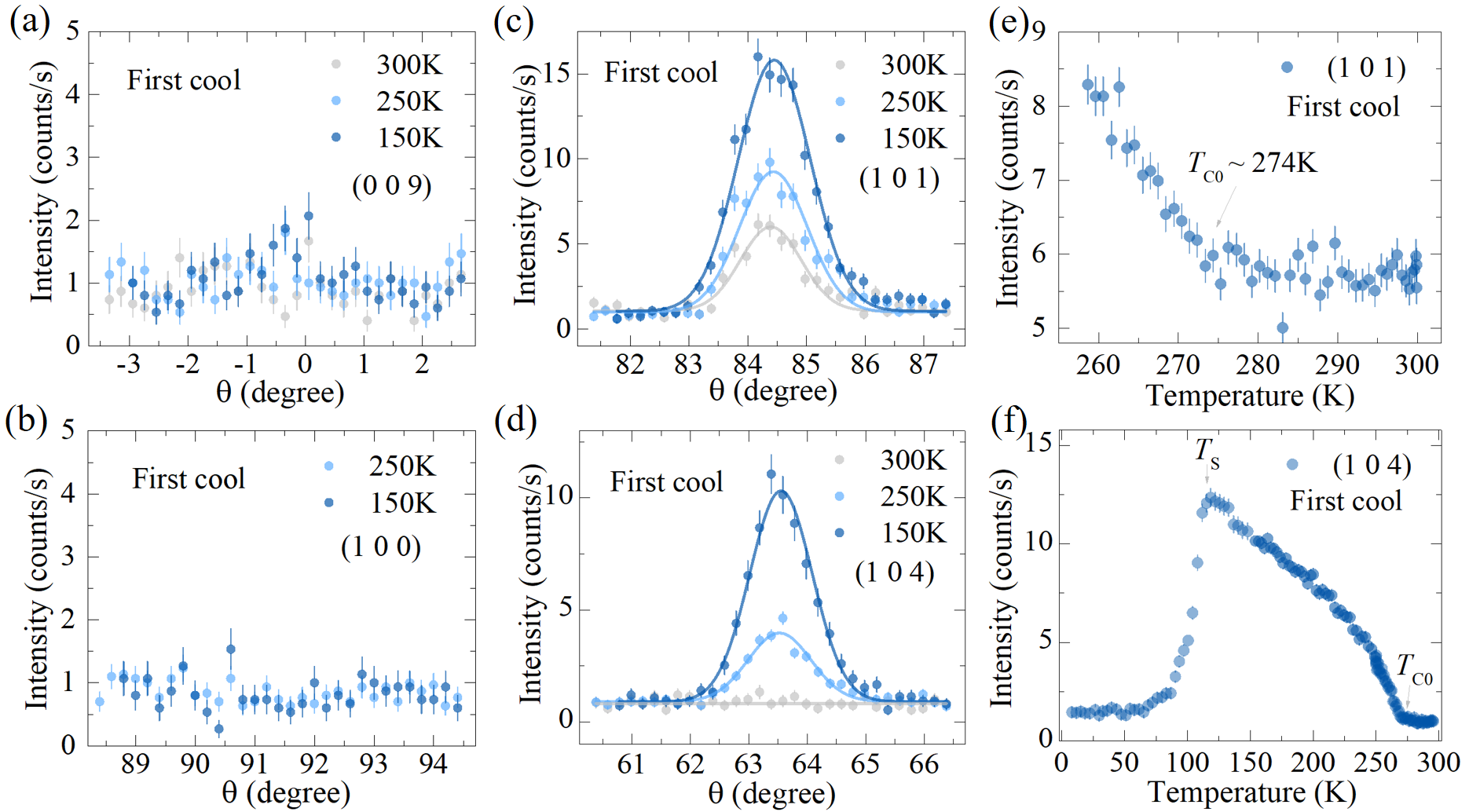}
\caption{(Color online) Temperature dependence of the neutron diffraction peaks of F5GT during the ``first cool" process. (a)-(d) Rocking scans at select Bragg peak positions: (a) (0, 0, 9), (b) (1, 0, 0), (c) (1, 0, 1) and (d) (1, 0, 4). Solid lines in (c)-(d) are Gaussian fits to the data. (e)-(f) Temperature dependence of the (1, 0, 1) peak (e) and (1, 0, 4) peak (f), with the same magnetic transition temperature $T_{\text{C0}}$ $\sim$ 274 K. The structural transition at $T_{\text{S}}$ is reflected by the sharp decrease of the (1, 0, 4) peak intensity in (f).}
\label{fig:Fig3_alpha}
\end{figure*}

\begin{figure*}[t]
\centering
\includegraphics*[width= 14.5 cm]{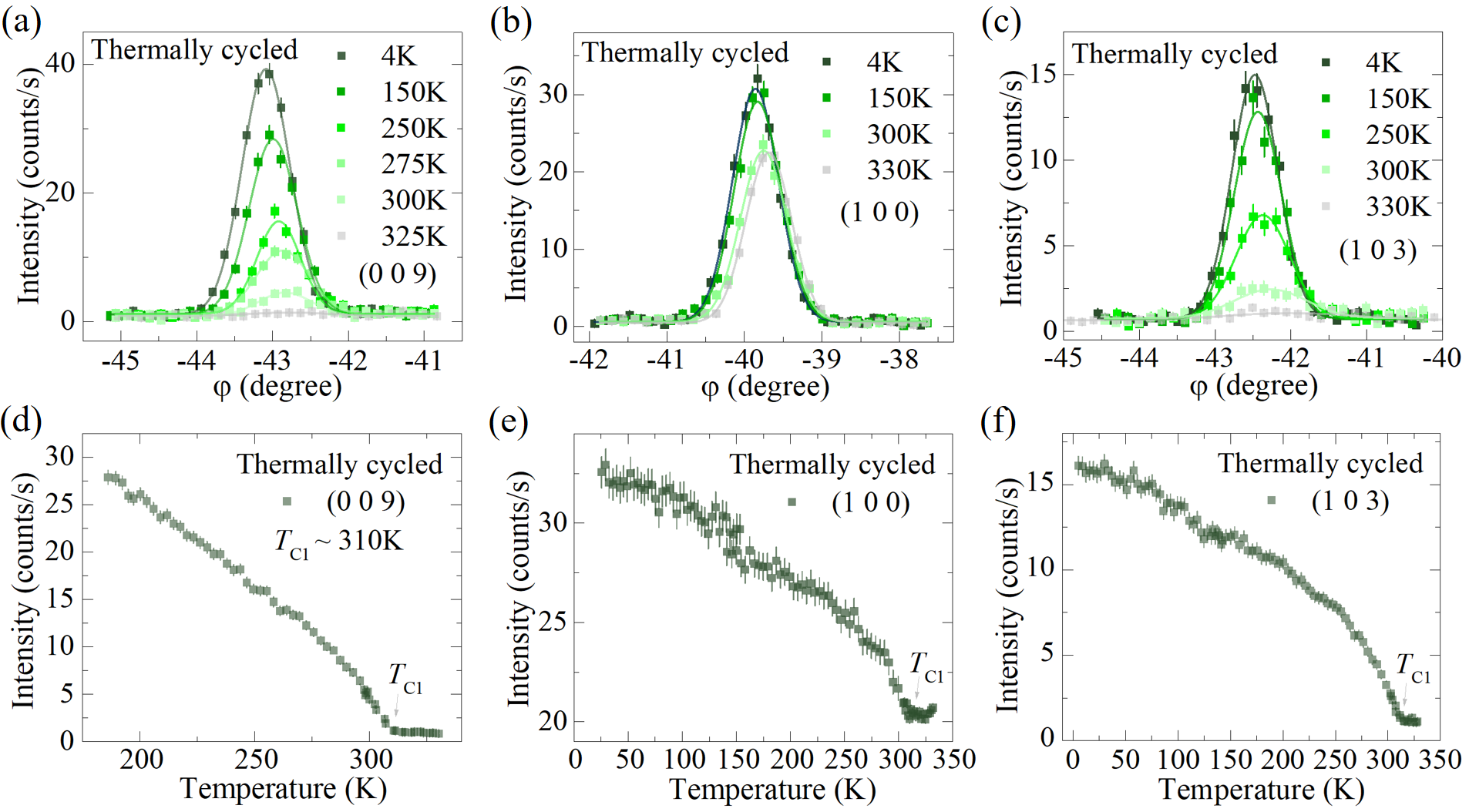}
\caption{(Color online) Temperature dependence of the neutron diffraction peaks of F5GT after the sample is ``thermally cycled". (a)-(c) Radial scans at select Bragg peak positions: (a) (0, 0, 9), (b) (1, 0, 0), (c) (1, 0, 3). Solid lines are Gaussian fits to the data. (d)-(f) Temperature dependence of the select Bragg peaks with an enhanced magnetic transition temperature $T_{\text{C1}}$ $\sim$ 310 K: (d) (0, 0, 9), (e) (1, 0, 0) and (f) (1, 0, 3).}
\label{fig:Fig4_alpha}
\end{figure*}

To understand the microscopic origin of the thermal history dependence of the magnetization, we firstly examine the structural properties of the quenched F5GT samples through single crystal x-ray diffraction (XRD), which presents minimum external perturbation to the sample. This is favored for characterizing the lattice structure of the vdW materials, because undesired stacking faults are frequently produced during the sample preparation for the commonly used powder XRD. A freshly quenched F5GT single crystal was subjected to an identical thermal treatment: the measurement was performed at RT first, gradually cooled down to $T$ = 10 K (``first cool") and then slowly warmed up (``first warm") towards RT. Fig. 2(a) shows the converted x-ray scattering data during the ``first cool" process. The observed peaks at RT can be well indexed with a space group (SG) R$\bar{3}$m (Fig. 1(a)) \cite{May_2019_ACSnano_F5GT}, indicated by the gray vertical lines and the labelled peaks. This is consistent with the rhombohedral layer stacking, labelled as ABC-stacking (Fig. 1(a)) \cite{May_2019_ACSnano_F5GT}. Interestingly, a new set of peaks emerge below $T_{\text{S}}$ $\sim$ 110 K during the ``first cool" process (Fig. 2). While the peaks associated with the ABC-stacking order are only weakly suppressed in magnitude with decreasing temperature, the intensities of the new peaks become stronger and stabilize at the lowest temperature, as indicated by the temperature dependent, normalized intensity of the selected peaks in Fig. 2(c). After the ``first cool" process, the diffraction pattern only displays a weak temperature dependence with respect to further thermal cycling, agreeing with the permanent change from the magnetization measurements \cite{May_2019_PRM}.

The new peaks evidenced from the x-ray scattering measurement, indicated and labelled in red in Fig. 2(a), can be well indexed with a SG P$\bar{3}$m1, signaling the formation of a closely related type of stacking order, $i.e.$, AA-stacking (Fig. 1(b)), as similarly observed in either Co or Ni doped F5GT samples \cite{May_2020_PRM_FCGT, Zhang_2021_FCGT, Chen_2022_PRL}. The two types of stacking order in F5GT, $i.e.$, ABC-stacking and AA-stacking, share the same motif while they only differ marginally through the stacking configuration along the lattice $c$ direction. In the AA-stacking order with primitive centering (Fig. 1(b)), the lattice $c$ of the unit cell should be 1/3 of the lattice $c$ with ABC-stacking. However, to maintain directly comparable Miller indices, a similar lattice $c$ $\approx$ 29.2 \AA \ is used for both AA-stacking and ABC-stacking orders here. The direct benefit is the clear distinction between the two types of stacking order in reciprocal space for the scattering measurement (Figs. 2-5). For ABC-stacking order, the (1, 0, $L$) type of nuclear Bragg peaks can only be observed when $L$ $\neq$ $3N$ ($L$ =  integers and $N$ = 0, 1, 2, ...). Meanwhile, for AA-stacking order indexed with this tripled lattice $c$, any allowed nuclear Bragg peak ($H$, $K$, $L$) must satisfy $L$ = 3$N$ ($N$ = 0, 1, 2, ...). Therefore, for (1, 0, $L$) type of nuclear Bragg peaks, the $\textbf{Q}_\text{AA}$ = (1, 0, 3$N$) ($N$ = 0, 1, 2, ...) positions will only be allowed in the AA-stacking ordered domains; while the $\textbf{Q}_\text{ABC}$ = (1, 0, 3$N$+1) or (1, 0, 3$N$+2) ($N$ = 0, 1, 2, ...) reflections will only be possible in the ABC-stacking ordered domains (Fig. 5(a)). A sharp distinction between the two types of stacking orders can be readily recognized by inspecting the (1, 0, $L$) type of structural peaks (Fig. 5(a)). Clearly, as shown in Fig. 2, the quenched F5GT only has the ABC-stacking order prior to the ``first cool" process because of the absence of the $\textbf{Q}_\text{AA}$ type of Bragg peaks. After the initial cooling below $T_{\text{S}}$, the emergence of the $\textbf{Q}_\text{AA}$ type of Bragg peaks, such as the (1, 0, 6) peak shown in Fig. 2(b), indicates the partial conversion from ABC-stacking to AA-stacking order, which permanently remains so hereafter.

\begin{figure}[t]
\centering
\includegraphics[width= 7.5 cm]{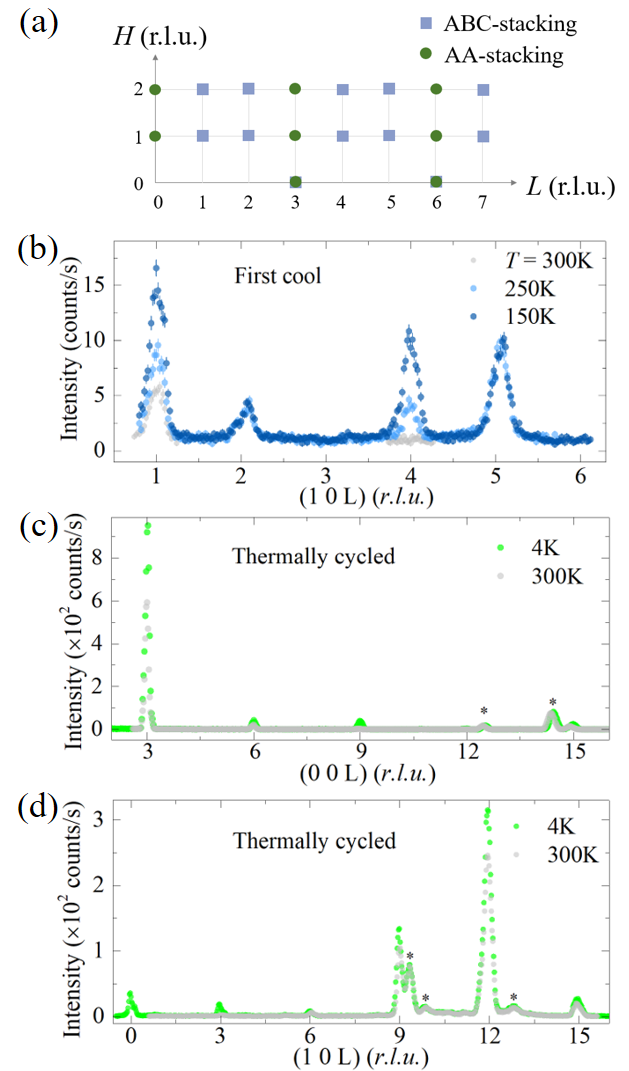}
\caption{(Color online) (a) Schematic of the nuclear Bragg peak positions in the ($H$, 0, $L$) plane of F5GT with two types of stacking order, indexed with the same lattice parameters: ABC-stacking (blue squares) and AA-stacking (green circles). (b)-(d) Long $L$ scans of the neutron diffraction peaks of F5GT at select temperatures, under different thermal status: (b) ``first cool" and (c)-(d) ``thermally cycled". Peaks marked by asterisks are temperature independent background from the aluminum holder.}
\label{fig:Fig5_alpha}
\end{figure}

The development of the AA-stacking order after the ``first cool" process through $T_{\text{S}}$ is also confirmed by our neutron diffraction study on another freshly quenched F5GT single crystal without any prior thermal treatment (Figs. 3-4 and Figs. 5(b)-(d)). The F5GT crystal was investigated following a similar temperature dependent sequence: starting at RT, ``first cool" to $T$ = 4 K and then warmed up to 350 K (``thermally cycled"). Without ``first cool" through $T_{\text{S}}$, the $\textbf{Q}_\text{AA}$ Bragg peaks, such as the (1, 0, 0), (1, 0, 3) and (1, 0, 6) reflections, are clearly absent (Fig. 3(b) and Fig. 5(b)). In agreement with the ABC-stacking order, the $\textbf{Q}_\text{ABC}$ Bragg peaks, such as the (1, 0, 1) and (1, 0, 4) reflections, are observed as expected (Figs. 3(c)-(d) and Fig. 5(b)). Interestingly, after the sample was subjected to ``first cool" to $T$ = 4 K, it permanently enters the ``thermally cycled" state with a different set of diffraction peaks. All of the $\textbf{Q}_\text{ABC}$ Bragg peaks observed before the structural transition at $T_{\text{S}}$, such as the (1, 0, 4) peak, completely vanish (Fig. 3(f) and Fig. 5(d)). Meanwhile, the $\textbf{Q}_\text{AA}$ Bragg peaks initially absent, such as the (1, 0, 0) peak, are now clearly visible (Figs. 4(b)-(c) and Fig. 5(d)). This suggests the complete conversion from ABC-stacking to AA-stacking order after the thermal cycling process in this particular sample. The combined x-ray and neutron scattering data on the structural properties evidently indicate the F5GT sample experiences a structural alteration at $T_{\text{S}}$ during the ``first cool" process, in which the system either partially or completely converts from ABC-stacking to AA-stacking order.

Having established this novel structural evolution in F5GT, the magnetic properties can also be re-evaluated from the neutron scattering data. Consistent with the early neutron powder diffraction study \cite{May_2019_ACSnano_F5GT}, the $\textbf{Q}_\text{ABC}$ Bragg peaks are both structural and magnetic in origin (Figs. 3(c)-(d) and Fig. 5(b)), confirming the FM nature of the magnetic order. For instance, both the (1, 0, 1) and (1, 0, 4) reflections, display temperature dependences with a magnetic onset temperature $T_{\text{C0}}$ $\sim$ 274 K (Figs. 3(e)-(f)), which is slightly lower than the value from the magnetization data of the other F5GT sample. Since the magnetic neutron scattering intensity is only sensitive to the spin moment component perpendicular to the scattering vector $\textbf{Q}$ \cite{Squires_1978_NeutronBook}, the absence of magnetic peaks at the (0, 0, 3$N$) ($N$ = 1, 2, ...) positions, such as at (0, 0, 9) in Fig. 3(a), confirms the out-of-plane spin moment direction in F5GT with ABC-stacking order. When this particular sample is ``thermally cycled" with the complete AA-stacking order, the magnetic peaks appear with a different pattern. Again, both the $\textbf{Q}_\text{AA}$ and (0, 0, 3$N$) reflections are structural and magnetic, and demonstrate the temperature dependence with an enhanced onset temperature $T_{\text{C1}}$ $\sim$ 310 K (Fig. 4 and Figs. 5(c)-(d)). For instance, the (0, 0, 9) peak now develops strong intensity below $T_{\text{C1}}$ (Figs. 4(a),(d)), in stark contrast to its absence before the ``first cool" treatment (Fig. 3(a)). The observation of magnetic peaks at the (0, 0, 3$N$) positions directly implies the existence of an in-plane spin moment component. Considering both the structural and magnetic peaks residing at the same Bragg peak positions, and from the representational analysis \cite{Wills2000}, the magnetic moment direction is either purely in-plane or out-of-plane. Therefore, the presence of the magnetic peaks at (0, 0, 3$N$) positions in the AA-stacking ordered F5GT sample implies FM order with the spin moments oriented in-plane. Additionally, a simple FM modelling based on the measured magnetic peak intensity with respect to the structural peak intensity results in an average spin moment size of $\sim$ 1.5(2) $\mu_{\text{B}}$/Fe. This value is slightly less than that obtained from the magnetization measurement ($\sim$ 2 $\mu_{\text{B}}$/Fe) \cite{Zhang_2021_FCGT, Chen_2022_PRL}, but close to estimated magnetic moments from other experimental probes and theoretical calculations \cite{May_2019_ACSnano_F5GT, Joe_2019_NMS, Ribeiro_2022_2DMA, Ershadrad_2022_JPCL}.

\section{Conclusion}

Our combined x-ray and neutron scattering study clarifies that the thermal history dependent magnetization of the quenched F5GT sample results from the structural alteration upon the ``first cool" procedure. Naturally, both the ABC-stacking and AA-stacking ordered domains coexist within the ``thermally cycled" sample. Importantly, because of the magneto-elastic coupling and the altered (interlayer) exchange couplings, the two types of domains demonstrate contrasting magnetic behavior. Specifically, the ABC-stacked domains host out-of-plane moments with a lower FM transition temperature $T_{\text{C0}}$; while the AA-stacked domains accommodate in-plane moments with an enhanced $T_{\text{C1}}$ $\sim$ 310 K (Figs. 1(a)-(b)). From this perspective, the enhanced $T_{\text{C}}$ and the anomalous critical component $\beta$ extracted from the magnetization data are well understood \cite{Zhang_2020_PRB_F5GT, Yamagami_2022_PRB}. This also partly reconciles the seemingly conflicting reports of the magnetic state in F5GT  \cite{May_2019_ACSnano_F5GT, May_2019_PRM, Chen_2022_2DM, Ribeiro_2022_2DMA, Zhang_2020_PRB_F5GT, Alahmed_2021_2DM, Schmitt_2022_skyrmionic_arxiv}. On the other hand, the AA-stacking order evidenced in F5GT is not surprising at all, since it has been documented in both Co or Ni doped F5GT systems \cite{May_2020_PRM_FCGT, Zhang_2021_FCGT, Zhang_2022_FCGT_sciadv, Chen_2022_PRL, Chen_2022_FeCo_PRM}. With higher magnetic onset temperatures consistently recorded also in Co/Ni doped FGT systems, the AA-stacking lattice structure seems to be favorable for increasing the magnetic ordering temperature.

It is worth noting that the ``thermally cycled" F5GT sample does not display a strong anomaly at $T_{\text{S}}$ from our neutron diffraction data (Figs. 4(e)-(f)), as opposed to the temperature induced spin-flip transition implied from the magneto-transport measurements \cite{Li_2022_AdvMat, Li_2021_arxiv}. Conversely, the sister compound Fe$_{4}$GeTe$_2$ behaves strikingly similar to pre-cooled F5GT in magnetization, including a similar $T_{\text{C}}$ $\sim$ 270 K, as well as a spin reorientation at $\sim$ 110 K \cite{Seo_2020_SciAdv_F4GT, Mondal_2021_PRB}. Therefore, the inconsistent magnetic behavior may be ascribed to the Fe deficiency $\delta$ in F5GT \cite{Seo_2020_SciAdv_F4GT, Liu_2022_CP}. Hence, it is crucial to characterize both the Fe deficiency level and the stacking order in F5GT \cite{Sivadas_2018_NanoLett}, especially when handling nanofalkes with a reduced sample thickness, in order to interpret correctly the magnetic properties.

In summary, our combined x-ray and neutron scattering study elucidates the thermal history dependence of the magnetization data of quenched F5GT single crystals. The irreversible, first order structural transition at $T_{\text{S}}$ $\sim$ 110 K through the ``first cool" process corresponds to the partial or complete alteration of the stacking order, from ABC-stacking to AA-stacking order. The neutron scattering data establish the FM nature and indicate that the magnetic properties, such as the spin moment direction and the magnetic onset temperature, are determined by the stacking order: the ABC-stacking ordered domains host out-of-plane spin moments with a lower $T_{\text{C0}}$ while the AA-stacking order presents in-plane spin moments with an increased $T_{\text{C1}}$ $\sim$ 310 K. Our work highlights the significant role of the stacking order of the lattice to the magnetism in F5GT, as well as the importance of the Fe deficiency in determining its magnetic ground state.

\bigbreak

\section{Acknowledgments}

Work at the University of California, Berkeley and the Lawrence Berkeley National Laboratory was funded by the U.S. Department of Energy, Office of Science, Office of Basic Energy Sciences, Materials Sciences and Engineering Division under Contract No. DE-AC02-05-CH11231 within the Quantum Materials Program (KC2202). This work is based upon research conducted at the Center for High Energy X-ray Sciences (CHEXS) which is supported by the National Science Foundation under award DMR-1829070. A portion of this research used resources at the High Flux Isotope Reactor, which is a DOE Office of Science User Facility operated by Oak Ridge National Laboratory.

\bibliography{Fe5GeTe2_main.bib}

\end{document}